\documentclass[11pt]{article}
\usepackage{moriond,epsfig}

\def\be{\begin{equation}}
\def\ee{\end{equation}}
\def\bea{\begin{eqnarray}}
\def\eea{\end{eqnarray}}

\newcommand     \epem           {\ifmmode{e^+e^-}\else{$e^+e^-$}\fi}
\def\beq{\begin{displaymath}}
\def\eeq{\end{displaymath}}

\def\beqn{\begin{eqnarray}}
\def\eeqn{\end{eqnarray}}
\def\ep{\epsilon}
\def\mayo{mayonnaise}

\begin{document}
\vspace*{3cm} 
\title{CHALLENGES IN THE CALCULATION OF NEXT-TO-NEXT-TO-LEADING ORDER
SCATTERING PROCESSES}


\author{ CARLO OLEARI }

\address{University of Wisconsin, Department of Physics \\
1150 University Avenue, Madison WI 53706, USA}

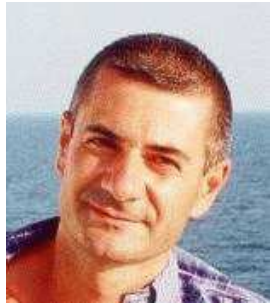
\maketitle\abstracts{
We discuss the status of the next-to-next-to-leading order
calculation in scattering processes, describing briefly the
challenges that have been overcome and what challenges are still to be
resolved.} 

\section{Introduction}
The Standard Model (SM), the quantum theory of the strong (QCD) and
electro-weak  interactions, is in very good shape.  In the last twenty
years, we have accumulated a wide body of experimental evidence of the
viability of the SM.
LEP and SLC, with a well understood initial state, have proven to be
ideal experiments for testing the strong and weak sector of the
theory. Systematic studies of large classes of jet shape variables have been
done to an extent that eliminates any doubt that perturbative QCD
is at work in $\epem$ annihilation.
HERA data on $e p$ collisions have shown clearly the scaling violation
and have given the best precise measurements of structure functions.
The highest energy scales currently available to test the SM are reached in
the hadron colliders.  These machines go beyond the task of testing: they are
discovery machines.  In fact, $W$ and $Z$ vector bosons and $t\bar{t}$ quark
cross sections were computed in a perturbative framework of the SM, and
represent remarkable examples where the model has allowed physicists to
predict cross sections for, at the time, unknown particles.

Even if the SM seems to work so well, there is an entire sector, the
Higgs boson sector, that has not yet been discovered.  We hope that LHC will
not only discover the Higgs boson, but will provide useful information about
the mass, couplings, charge, spin and color properties of the Higgs boson.
In addition to this not yet discovered sector, the SM has some ``disturbing''
features that lead us to think that it is not the final theory: questions
exist as to why there are three families of fermions and why there are so
many ad hoc parameters. In addition, more profound problems such as the
strong CP violation, the cosmological constant and the hierarchy problem have
not been completely understood.
Finally, a quantum theory of gravity has not yet been formulated,
so that the SM looks more like an effective theory, that should be integrated
or embraced by a more general theory.

Any extension of the SM has to face the fact that the SM has predicted with
high degree of accuracy the results from the experiments at high-energy
accelerators.
This means that any signal of new physics must be very small, and reaching a
higher level of precision in our understanding of hard production
phenomena becomes a fundamental issue in searching for new physics.
This goal can be achieved in two ways:\\
\indent -\ with better detectors and experimental analysis;\\
\indent -\ by refining our theoretical calculations.\\
One way (and surely not the only one) to refine our theoretical calculations is
to go from next-to-leading order (NLO)  to next-to-next-to-leading order
(NNLO).
There are several reasons for why this step is vital in reducing the
theoretical uncertainties:
\vspace{-0.3cm}
\begin{itemize}
\item[-] the dependence from the unphysical renormalization and factorization
scales is going to be reduced;
\vspace{-0.3cm}
\item[-] the presence of an additional hard parton gives rise to better
matching between the partonic and the hadronic final state;
\vspace{-0.3cm}
\item[-] double radiation from one of the incoming partons or single
radiation from both of the two incoming partons creates  more complicated 
transverse-momentum patterns for the final state partons, and may provide a
better and more theoretically motivated description of the data, without the
need of an intrinsic $k_T$ for the parton in the incoming hadron;
\vspace{-0.3cm}
\item[-] the need of power correction contributions to event shape variables
is going to be reduced, since part of the $1/Q$ contributions will be taken
into account by the NNLO term.
\end{itemize}

\section{The perfect \mayo}
As it is well known among the Italian chefs, in order to make \mayo, one must
have all the ingredients and then must follow the recipe.  But, even in this
case, one is not sure that all the ingredients will amalgamate correctly to
produce \mayo\ and will not instead create simple scrambled eggs.

\subsection{Ingredients}
Not all the ingredients are currently available. 
The matrix elements at NNLO order that are needed for $2\to 2$ 
scattering ( $\epem \to \epem$, $gg\to gg$, $qq' \to qq'$,
$q \bar{q} \to g \gamma $ \ldots) are the following:
\begin{enumerate}
\vspace{-0.3cm}
\item the square of $2\to4$  tree amplitude,
\vspace{-0.3cm}
\item the interference of the $2\to3$  one-loop amplitude
 with the tree-level one,
\vspace{-0.3cm}
\item the square of the $2\to2$ one-loop amplitude,
\vspace{-0.3cm}
\item the interference of the $2\to2$  two-loop amplitude
 with the  tree-level one,
\end{enumerate}
\vspace{-0.2cm}
where all the particles are taken massless and the external legs are
light-like.

{\em Mutatis mutandis}, these are the same ingredients that are needed in
$\epem \to  Z,\gamma  \to$ 3 jets, at LEP or at a future linear collider.
In fact, in this case, one of the incoming particles becomes a final one, and
the remaining incoming particle is time-like.  With a space-like incoming
particle, the same ingredients can describe deep-inelastic scattering in $ep$
colliders.

All these matrix elements are now available, mainly due to the removal of a
major stumbling block in the calculation of two-loop diagrams.

The first calculation of a two-loop four-point scattering amplitude was
performed in the case of the maximal-helicity-violating gluon-gluon
scattering.~\cite{BDK} Subsequently, generic $2 \to 2$ scattering matrix
elements at two loops have become tractable for massless particle exchanges
in the loops and with light-like external legs. Analytic expansion in
$\ep=(4-D)/2$, where $D$ is the space-time dimension, have been
computed~\cite{planarA,nonplanarA} and, at the same time, algorithms were
developed for the tensor reduction to master integrals of all relevant
two-loop topologies.~\cite{planarB}$^{-\,}$\cite{onshell6}

This technology has already been applied to a wide range of physically
interesting processes.  The interference of tree and two-loop graphs
(together with the simpler self interference of one-loop diagrams) for
various processes have now been computed, including Bhabha scattering
($e^+e^- \to e^+e^-$) in the massless electron limit~\cite{BDG} and all the
QCD $2 \to 2$ parton-parton scattering processes ($gg \to gg$, $gg \to q\bar
q$ and $q \bar q \to q\bar
q$).~\cite{qqQQ_qqqq_qqgg,gggg,1loopgggg}  Two-loop helicity
amplitudes have also been derived for gluon fusion into photons ($g g \to
\gamma \gamma$),~\cite{ggpp} light-by-light scattering ($\gamma \gamma \to
\gamma \gamma$)~\cite{pppp} and gluon-gluon scattering ($gg \to
gg$).~\cite{gggg_BDD}

The case where the internal propagators are massless but one external leg is
off-shell has also been intensively studied, leading to the evaluation of all
associated planar and non-planar master integrals~\cite{mi,smirnov_offshell}
needed in the NNLO computation of the decay of an off-shell photon to three
partons ($\gamma^* \to q\bar q g$), relevant for three jet production in
electron-positron annihilation.~\cite{zqqg}

\subsection{A missing ingredient}
In hadron-hadron collisions, factorization of the collinear singularities
from the incoming partons requires the evolution of the parton density
functions to be known to an accuracy matching the hard scattering matrix
element.  This entails knowledge of the three-loop splitting functions.
An approximate expression based upon the calculations of some moments in
Mellin space,~\cite{moms1,moms2} the knowledge
of the most singular $\log(1/x)$ behaviour at small $x$~\cite{Catani:1994sq}
and some exactly known terms~\cite{Gracey:1994nn}
has been computed by van Neerven \& Vogt.~\cite{vanNeerven:2000wp}
Very recently a set of NNLO parton-distribution functions, which was
obtained by fitting the new precise data for deep-inelastic scattering from
HERA and for inclusive jet production at the Tevatron, has been provided by
the MRST collaboration.~\cite{Martin:2002dr}

\subsection{The recipe}
The infrared singularities present in the two-loop $2 \to 2$ contributions
must cancel against the contributions from the one-loop $2 \to 3$ processes
when one particle is unresolved, and the contribution from the tree-level $2
\to 4$ processes when two particles are unresolved.  Unresolved particles are
either soft or collinear with one of the other partons in the event, and both
of these configurations have the appearance of a $ 2 \to 2$ scattering.
While the soft and collinear limits of these amplitudes has been vastly
investigated,~\cite{Gehrmann-DeRidder:1998gf}$^{-\,}$\cite{Kosower:1999rx}
a systematic procedure for analytically carrying through the integration of
the squared amplitudes over the soft and collinear phase space and the
cancellation of the $\ep$ poles has not yet been established.

\subsection{The final mixing}
A lot of thinking and work remains to be done.  In fact, even when all the
ingredients will be available and the entire recipe for the cancellation of
the infrared poles will be formulated, we still have the task of assembling
all the finite pieces in a fast and reliable partonic Monte Carlo program.

This task should not be underestimated, since speed and numerical accuracy
are main goals in this part of the process. This will be the most time
consuming part in the entire project,
since a lot of effort should be put in speeding up the program, helping the
Monte Carlo integration with importance sampling and phase-space remapping.

This is analogous to the whipping part of the \mayo-making process. In fact,
even if one puts together all the ingredients and follows the
recipe, mixing may not work.  It would be a pity if the \mayo\ would
not reach a fluffy and light consistency.

\section*{References}

\newcommand\hepph[1]{[{\tt hep-ph/#1}]}
\newcommand\hepth[1]{[{\tt hep-th/#1}]}
\newcommand\hepex[1]{[{\tt hep-ex/#1}]}

\relax

\def\pl#1#2#3{{\em Phys.\ Lett.\ }{\bf #1}, #2\ (#3)}
\def\np#1#2#3{{\em Nucl.\ Phys.\ }{\bf #1}, #2\ (#3)}
\def\jhep#1#2#3{{\em JHEP\ }{\bf #1}, #2\ (#3)}
\def\pr#1#2#3{{\em Phys.\ Rev.\ }{\bf #1}, #2\ (#3)}
\def\npps#1#2#3{{\em Nucl.\ Phys.\ Proc.\ Suppl.\ }{\bf #1}, #2\ (#3)}

\def\zp#1#2#3{{\it Z.\ Phys.\ }{\bf #1}\ (19#2)\ #3}
\def\prl#1#2#3{{\it Phys.\ Rev.\ Lett.\ }{\bf #1}\ (19#2)\ #3}
\def\rmp#1#2#3{{\it Rev.\ Mod.\ Phys.\ }{\bf#1}\ (19#2)\ #3}
\def\prep#1#2#3{{\it Phys.\ Rep.\ }{\bf #1}\ (19#2)\ #3}
\def\sjnp#1#2#3{{\it Sov.\ J.\ Nucl.\ Phys.\ }{\bf #1}\ (19#2)\ #3}
\def\app#1#2#3{{\it Acta Phys.\ Polon.\ }{\bf #1}\ (19#2)\ #3}
\def\jmp#1#2#3{{\it J.\ Math.\ Phys.\ }{\bf #1}\ (19#2)\ #3}
\def\jp#1#2#3{{\it J.\ Phys.\ }{\bf #1}\ (19#2)\ #3}
\def\nc#1#2#3{{\it Nuovo Cim.\ }{\bf #1}\ (19#2)\ #3}
\def\lnc#1#2#3{{\it Lett.\ Nuovo Cim.\ }{\bf #1}\ (19#2)\ #3}
\def\ptp#1#2#3{{\it Progr. Theor. Phys.\ }{\bf #1}\ (19#2)\ #3}
\def\tmf#1#2#3{{\it Teor.\ Mat.\ Fiz.\ }{\bf #1}\ (19#2)\ #3}
\def\tmp#1#2#3{{\it Theor.\ Math.\ Phys.\ }{\bf #1}\ (19#2)\ #3}
\def\epj#1#2#3{{\it Eur.\ Phys. J.\ }{\bf #1}\ (19#2)\ #3}
\relax


\begin{thebibliography}{99}

\bibitem{BDK}
Z.~Bern, L.~Dixon and D.A.~Kosower, 
\jhep{0001}{027}{2000} \hepph{0001001}.

\bibitem{planarA}
V.A.~Smirnov, 
\pl{B460}{397}{1999} \hepph{9905323}.

\bibitem{nonplanarA}
J.B.~Tausk, 
\pl{B469}{225}{1999} \hepph{9909506}.

\bibitem{planarB}
V.A.~Smirnov and O.L.~Veretin, 
\np{B566}{469}{2000}  \hepph{9907385}.

\bibitem{nonplanarB}
C.~Anastasiou, T.~Gehrmann, C.~Oleari, E.~Remiddi and J.B.~Tausk,  
\np{B580}{577}{2000} \hepph{0003261}.

\bibitem{AGO3}
C.~Anastasiou, E.W.N.~Glover and C.~Oleari,
\np{B575}{416}{2000},  Erratum-ibid.\ {\bf B585}, 763 (2000) \hepph{9912251}.

\bibitem{onshell5}
T.~Gehrmann and E.~Remiddi, 
\npps{89}{251}{2000} \hepph{0005232}.

\bibitem{onshell6}
C.~Anastasiou, J.B.~Tausk and M.E.~Tejeda-Yeomans, 
\npps{89}{262}{2000} \hepph{0005328}.

\bibitem{BDG}
Z.~Bern, L.~Dixon and A.~Ghinculov, 
\pr{D63}{053007}{2001} \hepph{0010075}.

\bibitem{qqQQ_qqqq_qqgg} 
C.~Anastasiou, E.W.N.~Glover, C.~Oleari and M.E.~Tejeda-Yeomans, 
\np{B601}{318}{2001} \hepph{0010212};
\np{B601}{341}{2000} \hepph{0011094}; 
\pl{B506}{59}{2001} \hepph{0012007}; 
\np{B605}{486}{2001} \hepph{0101304}. 

\bibitem{gggg} 
E.W.N.~Glover, C.~Oleari and M.E.~Tejeda-Yeomans, 
\np{605}{467}{2001} \hepph{0102201}.

\bibitem{1loopgggg} 
E.W.N.~Glover and M.E.~Tejeda-Yeomans, 
\jhep{0105}{010}{2001} \hepph{0104178}.

\bibitem{ggpp} 
Z.~Bern, A.~De Freitas and L.J.~Dixon, 
\jhep{0109}{037}{2001} \hepph{0109078}.
 
\bibitem{pppp} 
Z.~Bern, A.~De Freitas, L.J.~Dixon, A.~Ghinculov and H.L.~Wong, 
\jhep{0111}{031}{2001} \hepph{0109079}. 

\bibitem{gggg_BDD}
Z.~Bern, A.~De Freitas and L.J.~Dixon, 
\jhep{0203}{018}{2002} \hepph{0201161}.

\bibitem{mi}
T.~Gehrmann and E.~Remiddi, 
\np{B601}{248}{2001} \hepph{0008287};
\np{B601}{287}{2001} \hepph{0101124}.

\bibitem{smirnov_offshell}
V.A.~Smirnov, 
\pl{B491}{130}{2000} \hepph{0007032}; 
\pl{B500}{330}{2001} \hepph{0011056}.

\bibitem{zqqg} 
L.W.~Garland, T.~Gehrmann, E.W.N.~Glover, A.~Koukoutsakis and E.~Remiddi, 
\np{B627}{107}{2002} \hepph{0112081}.

\bibitem{moms1}  
S.A.~Larin, T.~van Ritbergen and J.A.M.~Vermaseren,
\np{B427}{41}{1994};
S.A.~Larin, P.~Nogueira, T.~van Ritbergen and J.A.M.~Vermaseren, 
\np{B492}{338}{1997} \hepph{9605317}.

\bibitem{moms2}  
A.~Retey and J.A.M.~Vermaseren, 
\np{B604}{281}{2001} \hepph{0007294}.

\bibitem{Catani:1994sq}
S.~Catani and F.~F.~Hautmann,
\np{B427}{475}{1994}.

\bibitem{Gracey:1994nn}
J.A.~Gracey,
\pl{B322}{141}{1994}.

\bibitem{vanNeerven:2000wp}
W.L.~van Neerven and A.~Vogt,
\pl{B490}{111}{2000}.

\bibitem{Martin:2002dr}
A.D.~Martin, R.G.~Roberts, W.J.~Stirling and R.S.~Thorne, 
\pl{B531}{216}{2002}.  


\bibitem{Gehrmann-DeRidder:1998gf}
A.~Gehrmann-De~Ridder and E.W.N.~Glover,
\np{B517}{269}{1998}.

\bibitem{Campbell:1998hg}
J.M.~Campbell and E.W.N.~Glover,
\np{B527}{264}{1998} \hepph{9710255}.

\bibitem{Catani:1998nv}
S.~Catani and M.~Grazzini,
\pl{B446}{143}{1999} \hepph{9810389}; 
\np{B570}{287}{2000} \hepph{9908523}; 
\np{B591}{435}{2000} \hepph{0007142}.

\bibitem{DelDuca:1999ha}
V.~Del~Duca, A.~Frizzo and F.~Maltoni,
\np{B568}{211}{2000} \hepph{9909464}.

\bibitem{Berends:1989zn}
F.A.~Berends and W.T.~Giele,
\np{B313}{595}{1989} .

\bibitem{Bern:1994zx}
Z.~Bern, L.J.~Dixon, D.C.~Dunbar and D.A.~Kosower,
\np{B425}{217}{1994}.

\bibitem{Bern:1998sc}
Z.~Bern, V.~Del~Duca and C.R.~Schmidt,
\pl{B445}{168}{1998} \hepph{9810409}.

\bibitem{Kosower:1999xi}
D.A.~Kosower,
\np{B552}{319}{1999} \hepph{9901201}.

\bibitem{Bern:1999ry}
Z.~Bern, V.~Del~Duca, W.B.~Kilgore and C.R.~Schmidt,
\pr{D60}{116001}{1999}.

\bibitem{Kosower:1999rx}
D.A.~Kosower and P.~Uwer,
\np{B563}{477}{1999} \hepph{9903515}.




\end{thebibliography}
\end{document}